\documentclass[twocolumn,secnumarabic,amssymb, nobibnotes, aps, prd]{revtex4-2}

\usepackage{graphicx}
\usepackage{float}
\usepackage{amsmath}
\usepackage{tikz}
\usepackage{cancel}

\newcommand{\Bi}{\mathrm{Bi}}
\newcommand{\Ma}{\mathrm{Ma}}
\newcommand{\e}{\mathrm{e}}

\graphicspath{}

\begin{document}

\title{Long-time emergent dynamics of liquid films undergoing thermocapillary instability}%

\author{Steven A. Kedda$^1$}
\author{Michael C. Dallaston$^1$} \email[Author contact: ]{michael.dallaston@qut.edu.au}

\author{Scott W. McCue$^1$}

\affiliation{$^1$School of Mathematical Sciences, Queensland University of Technology, Brisbane, QLD, 4001, Australia.}

\begin{abstract}
The study of viscous thin film flow has led to the development of highly nonlinear partial differential equations that model how the evolution of the film height is affected by different forces. We investigate a model of interaction between surface tension and the thermocapillary Marangoni effect, with a particular focus on the long-time limit.  
In this limit, the model predicts the creation of an infinite cascade of successively smaller satellite droplets near points where the film thickness vanishes. 
Motivated by recent progress on the analysis of discrete self-similarity in thin film equations, we compute solutions in a space- and time-rescaled coordinate system.  Using this rescaled system we observe the dynamics much further in time than has previously been achieved.  
The observed behaviour is close to, but distinct from, previous observations of discretely self-similar thin film flows, in that the rescaled system does not settle down to a periodic solution, but instead has aspects that continue to evolve monotonically in scaled time.  This discovery suggests there are as-yet unexplored ways in which discrete self-similarity may be exhibited.

\end{abstract}

\maketitle

\section{Introduction}

\begin{figure}
\centering
%
%
%
%
%
%
%
\includegraphics{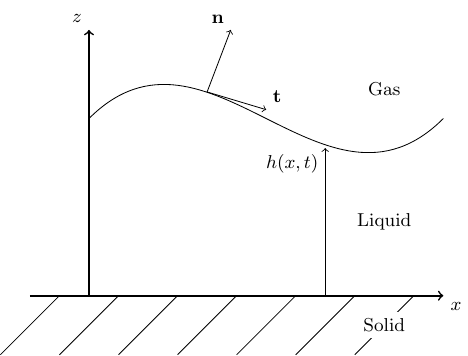}
\caption{Schematic of the thin film model, with local film height $h(x,t)$, normal $\textbf{n}$, and tangent $\textbf{t}$. \label{fig:thinfilm1}
}
\end{figure}

The complex emergent behavior of solutions to nonlinear partial differential equations is a central topic in applied mathematics and physics.  In fluid mechanics, the flow of a viscous thin film on a substrate is often modelled with a single evolution equation for the local film thickness $h(x,t)$ as a function of space $x$ and time $t$\,\cite{o2002theory, myers1998thin, craster_dynamics_2009, eggers_nonlinear_1997} (see Fig.~\ref{fig:thinfilm1}).  Under idealized or approximate circumstances, this equation is frequently of the form
\begin{equation}
\frac{\partial h}{\partial t} + \frac{\partial}{\partial x}\left(h^m \frac{\partial^3 h}{\partial x^3} + h^n \frac{\partial h}{\partial x}\right) = 0,
\label{eq:Thin_film_eqn}
\end{equation}
where the exponents $m$ and $n$ depend on the geometry and the physical forces under consideration;  the fourth-order spatial term in \eqref{eq:Thin_film_eqn} arises due to surface tension, while the second-order term may arise from any number of destabilising effects \cite{craster_dynamics_2009,oron_long-scale_1997,myers1998thin}.

The nonlinear behavior of \eqref{eq:Thin_film_eqn} is strongly dependent on the values of the exponents $m$ and $n$.  For example, equation \eqref{eq:Thin_film_eqn} with $m=3$ and $n=-1$ is used to model finite-time thin film rupture due to attractive van der Waals forces.  Mathematically, such a model exhibits classical self-similarity, in which the thickness near the rupture point asymptotically tends to a profile that is unchanging under an appropriate time-dependent rescaling of the spatial dimensions~\cite{witelski_stability_1999,zhang_similarity_1999}.  Recently it has been shown that for different values of the exponents $m$ and $n$, \eqref{eq:Thin_film_eqn} may instead exhibit \textit{discrete} self-similarity, in which the profiles are only self-similar at discretely chosen times \cite{dallaston_self-similar_2017,dallaston_discrete_2018}.  The boundaries between different behaviors in $(m,n)$ parameter space has recently been explored in Ref.~\cite{dallaston_regular_2021}.  A more general description of discrete and classical self-similarity is provided in Refs \cite{sornette1998discrete,eggers_singularities_2015}.

The similarity analysis described above has thus far concentrated on equations that exhibit finite-time rupture, meaning $h\to 0$ as time $t$ approaches a finite value.  However, a particularly interesting application of \eqref{eq:Thin_film_eqn} is destabilisation, thinning, and pattern formation of a heated thin film due to thermocapillary (Marangoni) stress,  for which \eqref{eq:Thin_film_eqn} with $m=3$ and $n=2$ represents an ideal case.  Using thermocapillarity (the dependence of surface tension on temperature) to control interfaces is of interest as a method of creating patterned surfaces\,\cite{singer2017thermocapillary}.
It has been observed for some time that the nature of thermocapillary thin film solutions is for the minimum thickness to tend to zero as time increases without bound (or infinite-time rupture), producing a cascade of structures of successively smaller sizes as it does so.  This property has been observed both in lubrication equations of the form \eqref{eq:Thin_film_eqn} and generalisations\,\cite{yeo2003marangoni,oron_long-scale_1997,burelbach_nonlinear_1988,shklyaev2012long,oron_nonlinear_2000}, as well as more comprehensive Stokes flow models that do not use the lubrication approximation \cite{boos_cascade_1999}.
In  Ref. \cite{shklyaev_superexponential_2010}, these structures are referred to as dissipative \textit{compactons}, as they represent (in the limit that time increases without bound) compactly supported steady-state solutions to \eqref{eq:Thin_film_eqn}.  As they are only asymptotically compactly supported, we will refer to these as satellite droplets in our study, in analogy to the structures that form in thread break-up\,\cite{brenner_iterated_1994,dallaston_stability_2021}.

While this cascade of structures present in models for thermocapillary flows is highly suggestive of the presence of discrete self-similarity, the analysis of thermocapillary thin film flows in such a framework has not previously been undertaken.  As the thinning process occurs over a considerably long time scale, numerical simulation has previously only been able to produce 3--4 generations of satellite droplets at most \cite{shklyaev_superexponential_2010}.  

In this article, we use a dynamically rescaled version of \eqref{eq:Thin_film_eqn}, motivated by the search for discrete self-similar behavior, to study the solution for very long times.  
In Section \ref{sec:formulation}, we summarize how modelling thin viscous films with the thermocapillary Marangoni effect leads to an evolution equation of the form \eqref{eq:Thin_film_eqn}.  In Section \ref{sec:numerics}, we describe the numerical solution of the thin film equation in which the initial formation of satellite droplets is observed, and which is necessary for choosing an appropriate initial condition for the dynamically rescaled equation.  In Section \ref{sec:rescalednumerics}, we show how a dynamically rescaled version of the evolution equation allows us to continue much further in time than has previously been possible.  Our rescaling is such that classical or discrete self similarity, if present, would correspond to a steady state or periodic solution, respectively.  However, we observe that evolution to a periodic solution occurs very slowly, if at all.  On the time interval we calculate over, each satellite droplet forms at a size slightly less than a constant fraction of the previous one (that is, at a slightly super-geometric rate), although the ratio of successive sizes does seem to slowly approach a constant value.  

\section{Problem Formulation}
\label{sec:formulation}

The following derivation uses methodology from the works in Ref. \cite{oron_long-scale_1997} in order to include the thermocapillary effect for free surface Stokes flow. We begin the derivation of our model by stating the Navier-Stokes equations in the fluid region, including an energy equation for temperature
\begin{align}
    \rho \left( \frac{\partial \textbf{u} }{\partial t} + (\textbf{u} \cdot \nabla)\textbf{u} \right)
    &= -\nabla p + \mu \nabla^2 \textbf{u}, \label{eq:gov1} \\
    \nabla \cdot \textbf{u} &= 0,\label{eq:gov2}\\
\rho c \left( \frac{\partial \vartheta }{\partial t} + (\textbf{u} \cdot \nabla) \vartheta \right)
    &= k_{th} \nabla^2 \vartheta,\label{eq:gov3}
\end{align}
with fluid velocity $\textbf{u} = (u,w)$, density $\rho$, viscosity $\mu$, pressure $p$, temperature $\vartheta$, specific heat $c$, and thermal conductivity $k_{th}$. Boundary conditions on the substrate $z=0$ are no-slip $u=w=0$ and fixed temperature $\vartheta = \vartheta_0$. The boundary conditions on the free surface $z=h$ are 
\begin{align}
\textbf{T} \cdot \hat{\textbf{n}} &= - \kappa \sigma \hat{\textbf{n}} + \dfrac{\partial \sigma}{\partial s} \hat{\textbf{t}},\\
w &= \dfrac{\partial h}{\partial t} + u \dfrac{\partial h}{\partial x},\\
k_{th} \nabla \vartheta \cdot \hat{\textbf{n}} &= \alpha_{th} \left(\vartheta_\infty - \vartheta \right),
\end{align}
describing the effects of interfacial shear and normal stresses, the kinematic boundary condition, and Newton's~law of cooling, respectively. Here we have the stress tensor $\textbf{T}$, outward normal unit vector $\hat{\textbf{n}}$, tangent unit vector $\hat{\textbf{t}}$, mean curvature $\kappa$, surface tension $\sigma$, interfacial arc length $s$, heat transfer coefficient $\alpha_{th}$, and ambient temperature $\vartheta_\infty$. The unit normal, unit tangent, and mean curvature are defined as
\begin{equation*}
\hat{\textbf{n}} = \frac{\left(-\frac{\partial h}{\partial x},1 \right)}{\sqrt{1 + \left(\frac{\partial h}{\partial x}\right)^2}}, \ \ \hat{\textbf{t}} = \frac{\left(1, \frac{\partial h}{\partial x}\right)}{\sqrt{1 + \left(\frac{\partial h}{\partial x}\right)^2}}, \ \ \kappa = \frac{\frac{\partial^2 h}{\partial x^2}}{\left(1 + \left(\frac{\partial h}{\partial x}\right)^2\right)^{3/2}}.
\end{equation*}
Surface tension is assumed to change linearly with respect to temperature (see eg. \cite{oron_long-scale_1997}), modelled by the following constitutive relation
\begin{equation}
\sigma(\vartheta) = \sigma_0 \left( 1 - \gamma \frac{\vartheta - \vartheta_\infty}{\vartheta_0 - \vartheta_\infty} \right), \nonumber
\end{equation}
with reference surface tension $\sigma_0 = \sigma(\vartheta_0)$, and nondimensional temperature dependence coefficient $\gamma$.

We nondimensionalize the governing equations and boundary conditions by introducing the following non-dimensional variables
\begin{align*}
\hat{x} &= x/L, \ \hat{z} = z/H, \ \hat{h} = h/H, \ \hat{t} = t/T,\\
\hat{u} &= u/U, \ \hat{w} = w/W, \ \hat{p} = p/P, \ \hat{\vartheta} = \frac{\vartheta - \vartheta_\infty}{\vartheta_0 - \vartheta_\infty},
\end{align*}
with the following choices for the dimensional scales $T=L/U$, $W=UH/L$, $P = L \mu U/H^2$, and $U = 3\sigma_0 \epsilon^3/\mu$.  These scales introduce the lubrication parameter $\epsilon$, Biot number $\Bi$, and Marangoni number $\Ma$, defined as
\begin{align*}
\epsilon = \frac{H}{L}, \quad \Bi = \frac{\alpha_{th} H}{k_{th}}, \quad 
\Ma = \epsilon \frac{\sigma_0 \gamma}{\mu U}.
\end{align*}
Substituting the non-dimensional parameters into the governing equations \eqref{eq:gov1} - \eqref{eq:gov3}, dropping the hat notation and applying the lubrication limit $\epsilon \ll 1$, we have governing equations
\begin{align}
\frac{\partial p}{\partial x} &= \frac{\partial^2 u}{\partial z^2}, \label{eq:horizontal_velocity}\\
\frac{\partial p}{\partial z} &= 0, \\
\frac{\partial u}{\partial x} + \frac{\partial w}{\partial z} &= 0, \label{eq:div_free}\\
\frac{\partial^2 \vartheta}{\partial z^2} &= 0, \label{eq:tempgov}
\end{align}
and boundary conditions on $z=h$
\begin{align}
p &= 3\frac{\partial^2 h}{\partial x^2},\\
\frac{\partial u}{\partial z} &= -\Ma\frac{\partial \vartheta}{\partial x}, \label{eq:shearBC}\\
w &= \frac{\partial h}{\partial t} + u \frac{\partial h}{\partial x}, \label{eq:kinematic}\\
\frac{\partial \vartheta}{\partial z} + \Bi \vartheta &= 0. \label{eq:tempBC}
\end{align}
The non-dimensional boundary conditions on the substrate $z=0$ are $u=w=0$ and $\vartheta = 1$.

We now derive a non-dimensional expression for the temperature through the height of the film. We integrate the simplified governing equation \eqref{eq:tempgov} twice in the $z$ direction and apply the boundary conditions $\vartheta(0) = 1$ and equation \eqref{eq:tempBC} to arrive at
\begin{align}
\vartheta = 1 - \frac{\Bi z}{1 + \Bi h}.
\end{align}
We derive an expression for the horizontal velocity profile by integrating \eqref{eq:horizontal_velocity} twice through $z$. With the no-slip boundary condition at $z=0$ and the shear stress condition \eqref{eq:shearBC}, we have
\begin{align}
u = \frac{\partial p}{\partial x} \left( \frac{z^2}{2} - hz \right) - \Ma \frac{\partial \vartheta}{\partial x} z. \label{eq:u}
\end{align}
Using the zero divergence condition \eqref{eq:div_free} and integrating through the height of the film, we find an expression for $w$ at the interface which is used in the kinematic boundary condition. Substituting these results for $u$ and $w$ into the kinematic boundary condition returns the evolution equation for the interface:
\begin{equation*}
\frac{\partial h}{\partial t} = \frac{\partial}{\partial x} \left( \frac{h^3}{3} \frac{\partial p}{\partial x} + \Ma\frac{h^2}{2} \frac{\partial \vartheta}{\partial x} \right).
\end{equation*}
Hence we have our governing PDE
\begin{equation}
\frac{\partial h}{\partial t} + \frac{\partial}{\partial x}\left(h^3 \frac{\partial^3 h}{\partial x^3} + \frac{\Ma\,\Bi h^2}{2(1+\Bi\, h)^2}\frac{\partial h}{\partial x}\right) = 0,
\label{eq:pdeWithParams}
\end{equation}
which is equivalent to equation (2.62) found in Ref. \cite{oron_long-scale_1997}.

In the nondimensionalisation thus far, no specific choice was made for the horizontal or vertical length scales, $L$ and $H$.  Generally, these are set by the initial condition and domain size.  However, since we ultimately focus on the behavior over a long time period, and in a vanishing spatial region, it is more appropriate to select length scales based on the dynamics.  We thus rescale \eqref{eq:pdeWithParams} by
\begin{equation*}
h \mapsto \left(\tfrac{1}{2}\Ma\,\Bi\right)^{-2}h, \qquad x \mapsto \left(\tfrac{1}{2}\Ma\,\Bi\right)^{-3/2}x.
\end{equation*}
In addition, as $h\to 0$ the term $(1+\Bi\, h)^2$ tends to unity.  For this reason the long-time dynamics of \eqref{eq:pdeWithParams} is described by the parameter-free equation
\begin{equation}
\frac{\partial h}{\partial t} + \frac{\partial}{\partial x}\left(h^3\frac{\partial^3 h}{\partial x^3} + h^2\frac{\partial h}{\partial x}\right) = 0,
\label{eq:MPDE}
\end{equation}
which is in the form considered in \cite{shklyaev_superexponential_2010}.

\section{Numerical simulation}
\label{sec:numerics}

In this section we describe numerical simulation of the equation \eqref{eq:MPDE}, exhibiting the initial dynamics that we will then continue using the numerical method described in Section \ref{sec:rescalednumerics} for longer times.  We use a central finite difference scheme on a uniform spatial grid, solved in time by using Matlab's ODE15s stiff ODE system solver.  Since \eqref{eq:MPDE} is in conservative form $h_t + Q_x = 0$, we discretize the flux $Q = h^3 h_{xxx} + h^2 h_x$ with a central finite difference in $h$, then apply a central finite difference to calculate its derivative, which results in a seven-point stencil. This two-step spatial discretisation appears to result in a more numerically stable scheme compared to expanding and directly discretising each individual term in the flux.\\

\begin{figure}
\centering
\includegraphics[scale=1]{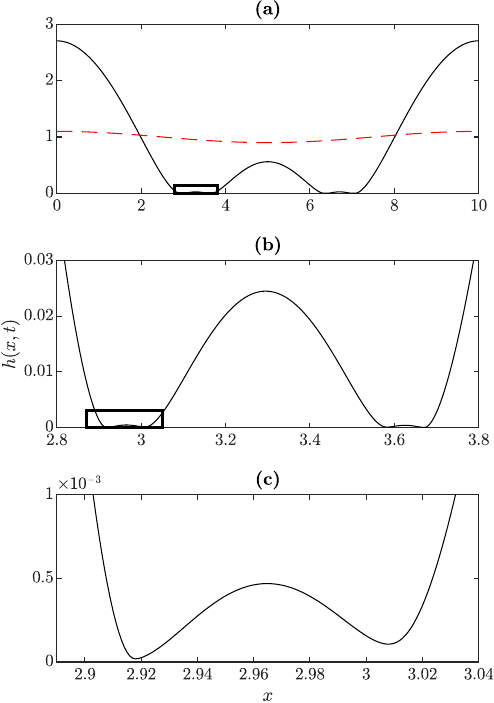}
\caption{Numerical solution to \eqref{eq:MPDE}-\eqref{eq:MPDE_IC} on periodic boundaries at $t=4000$, with choices $\delta = 0.1$ and $L = 10$. (a) The solution on the entire computational domain in black online with initial condition in red dashed line, with (b) and (c) magnification of the indicated regions, revealing the continuing formation of satellite droplets. \label{fig:marangoni2}}
\end{figure}

Using the above method we solve \eqref{eq:MPDE} on a domain $x \in [0,L]$ with periodic boundary conditions, and with an initial condition comprising a perturbed flat interface:
\begin{equation}
h(x,0) = 1 + \delta\cos\left(\frac{2\pi x}{L}\right). \label{eq:MPDE_IC}
\end{equation}
In Fig.~\ref{fig:marangoni2} we show the result of this simulation with $\delta = 0.1$, $L = 10$.  The flat interface is unstable for this wavelength perturbation (as is easily demonstrated by linear stability analysis) and the initial perturbation grows in amplitude.  After some time the perturbation bifurcates into two local minima.  Subsequently, the film thickness tends toward zero at two points as time increases.  Between these two points the remaining fluid forms a bump, referred to as a compacton in Ref.~\cite{shklyaev_superexponential_2010}, and a satellite droplet in the present study.  As the film thins further, each local minimum bifurcates into two minima, resulting in a further droplet on a smaller scale.
 With $N=10000$ spatial nodes we can accurately simulate this equation up to roughly $t=4000$, beyond which the spatial accuracy required near rupture becomes too challenging, making it extremely difficult to observe further generations of satellite droplets when solving \eqref{eq:MPDE} directly.
 
\begin{figure}
\centering
\includegraphics[scale=1]{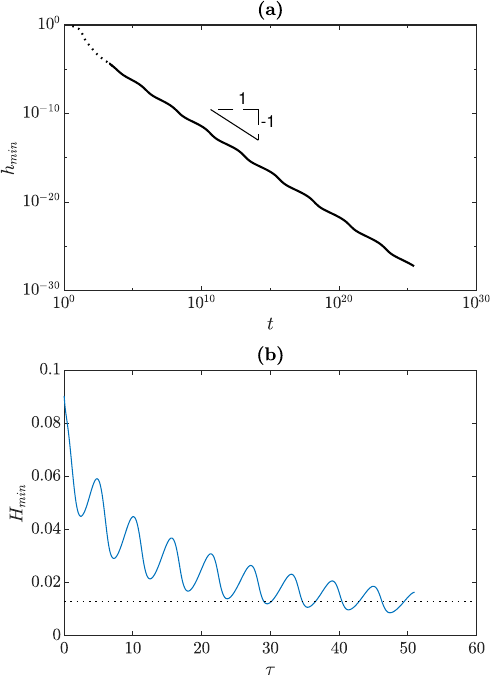}
\caption{(a) The minimum film thickness $h_\mathrm{min}(t)$ over time, on a logarithmic scale. The dotted line is the solution to the unscaled problem \eqref{eq:MPDE}-\eqref{eq:MPDE_IC}, while the solid line comes from the solution to the dynamically rescaled problem \eqref{eq:sMPDE}-\eqref{eq:sMPDE_BC} transformed back to the original variables $h$ and $t$ for comparison. The minimum thickness tends to a power law proportional to $t^{-1}$, but oscillates indefinitely around this power law. (b) the scaled minimum thickness $H_{\min}(\tau)$, from which the bulk of the minimum thickness in (a) is computed, is observed to oscillate around a value that is slowly approaching $\sim 0.1$.\label{fig:marangoni_loglog}}
\end{figure}

In Fig.~\ref{fig:marangoni_loglog} we plot the minimum thickness $h_\mathrm{min}(t)$ as a function of time $t$; the dotted portion of the curve represents the results of the simulation performed in this section.  If the solution were to be self-similar in this time span, we would expect the minimum thickness to approach a power law (equivalent to a straight line on logarithmic axes).  However, for the time scales achievable when solving \eqref{eq:MPDE} directly, no convergence to a power law is observed.  We are thus motivated to solve a dynamically rescaled form of  \eqref{eq:MPDE} that actively focuses on the region where the thickness goes to zero, and will allow us to observe solution behavior for much larger times.

\section{Rescaled equation for long time behavior}
\label{sec:rescalednumerics}

In order to observe the behavior of solutions to \eqref{eq:MPDE} on long time scales, and close to a point at which the film thickness vanishes, we apply the following change of variables:
\begin{equation}
h(x,t) = t^{-1} H(\eta, \tau), \quad \eta = \frac{x-x_0}{t^{-1/2}}, \quad \tau = \ln t,
\label{eq:scale}
\end{equation}
to \eqref{eq:MPDE}, resulting in the new equation for $H(\eta, \tau)$
\begin{equation}
\frac{\partial H}{\partial \tau} =  H - \frac{1}{2} \eta \frac{\partial H}{\partial \eta} - \frac{\partial}{\partial \eta} \left[ H^3 \frac{\partial^3 H}{\partial \eta^3} + H^2 \frac{\partial H}{\partial \eta} \right]. \label{eq:sMPDE}
\end{equation}
Here $x_0$ is an assumed rupture point, that is, where $h(x_0,t) \to 0$ as $t\to\infty$.

As any fixed value of $x \neq x_0$ corresponds to $\eta \to \pm\infty$ as $t \rightarrow \infty$, \eqref{eq:sMPDE} must be posed on the infinite domain $-\infty < \eta < \infty$.  Since we expect the time evolution of the film thickness to go to zero away from $x_0$ much faster than at $x_0$, appropriate far field conditions are
\begin{equation}
\frac{\partial H}{\partial \tau} \sim H - \frac{1}{2} \eta \frac{\partial H}{\partial \eta}, \ \eta \rightarrow \pm \infty, \label{eq:sMPDE_BC}
\end{equation}
used by Ref.~\cite{dallaston_discrete_2018} for finite-time rupture.

The substitution \eqref{eq:scale} is inspired by, but not limited to, the search for self-similar behavior \cite{witelski_stability_1999}.  A classical similarity solution of \eqref{eq:MPDE} corresponds to a steady state of \eqref{eq:sMPDE}, which is known to exist and be stable for equations similar to \eqref{eq:Thin_film_eqn} with certain exponents, such as for models of thin film rupture due to van der Waals force \cite{zhang_similarity_1999,witelski_stability_1999}.  In our case, a similarity solution would satisfy the boundary value problem
\begin{equation}
0 = H - \frac{1}{2} \eta H' - \left[ H^3 H''' + H^2 H' \right]'
\label{eq:sMPDE_steady}
\end{equation}
\begin{equation}
H \sim C \eta^2, \qquad \eta \rightarrow \pm \infty
\label{eq:steadybcs},
\end{equation}
for constant $C$, where the far-field algebraic behavior \eqref{eq:steadybcs} results from the steady version of \eqref{eq:sMPDE_BC}.  However, we have been unable to find a solution to \eqref{eq:sMPDE_steady}--\eqref{eq:steadybcs} by applying the method in Ref.~\cite{witelski_stability_1999} (the similarity solutions reported in Ref.~\cite{shklyaev_superexponential_2010} do not satisfy the appropriate bounday condition \eqref{eq:steadybcs}, which means they do not correspond to behavior exhibited by \eqref{eq:MPDE}).  This apparent absence of a similarity solution is consistent with the observations of the long-time behavior of the time-dependent problem \eqref{eq:sMPDE}-\eqref{eq:sMPDE_BC} that we make shortly.

The inclusion of dependence on logarithmic time $\tau$ in \eqref{eq:sMPDE}-\eqref{eq:sMPDE_BC} means that non-classical self-similar behavior can also be studied; recent work has shown how equations of the form \eqref{eq:Thin_film_eqn} can (depending on the value of the exponents) exhibit discrete, rather than classical, self-similar behavior, corresponding to periodic (in $\tau$) solutions to their counterparts to \eqref{eq:sMPDE}. Such discrete similarity solutions are of interest here as they form an infinite cascade of structures of geometrically decreasing size, similar to the cascade of satellite droplets exhibited by \eqref{eq:MPDE}.  More generally, however, \eqref{eq:sMPDE} may be solved numerically without making any prior assumption on its eventual dynamics, although accurate estimation of the rupture location $x_0$ is crucial, as we describe further below.

To simulate \eqref{eq:sMPDE}, we need to create a suitable initial condition.  By taking a late-time solution to the problem in the original coordinate system \eqref{eq:MPDE}, and scaling it into the new coordinate system $(\eta, H)$, we may view the solution as a continuation of the original one for long times.  From solving the unscaled problem to $t = 2000$, where we still have relatively high resolution, we use the scaling equations \eqref{eq:scale} to convert the numerical solution $h(x,t=2000)$ to an initial condition $H(\eta,0)$ for \eqref{eq:sMPDE}.  

In order to carry out this conversion, the infinite-time rupture location $x_0$ must be numerically estimated, as it cannot be determined exactly through symmetry arguments.  Indeed, the arbitrary value of $x_0$ is due to the spatial invariance of the original equation \eqref{eq:MPDE}, which corresponds to an artificial instability in the scaled problem \eqref{eq:sMPDE} \cite{dallaston_discrete_2018,witelski_stability_1999}.  In selecting an initial condition for $H$, the value of $x_0$ must be carefully chosen to avoid the minimum film height rapidly leaving the computational window.  We use a bisection approach to estimate $x_0$ by initially choosing the $x$ coordinate of the minimum film height, solving the scaled problem until the minimum film height moves away from the neighborhood of $\eta = 0$, then adjusting the estimate of $x_0$ so the scaled solution retains its minimum near $\eta = 0$ for as long as possible.  

We solve equation  \eqref{eq:sMPDE} numerically, using a finite difference method similar to that described in Section \ref{sec:numerics}.  The equation is solved on the domain $\eta \in [-2.5, 7.5]$ with 600 spatial nodes, with the far-field condition \eqref{eq:sMPDE_BC} implemented using upwind approximations of the derivatives at the last 3 spatial nodes at each end.  With a good quality unscaled solution as the initial condition and $x_0$ chosen carefully through the method described above, the solution may be continued to values $\tau \approx 50$, corresponding to original time $t = O\left(\mathrm{e}^{50}\right)$.  This length of time is not possible to achieve in the original coordinate system, and is far beyond that reached previously \cite{shklyaev_superexponential_2010}.

\begin{figure*}
\centering
\includegraphics[scale=.8]{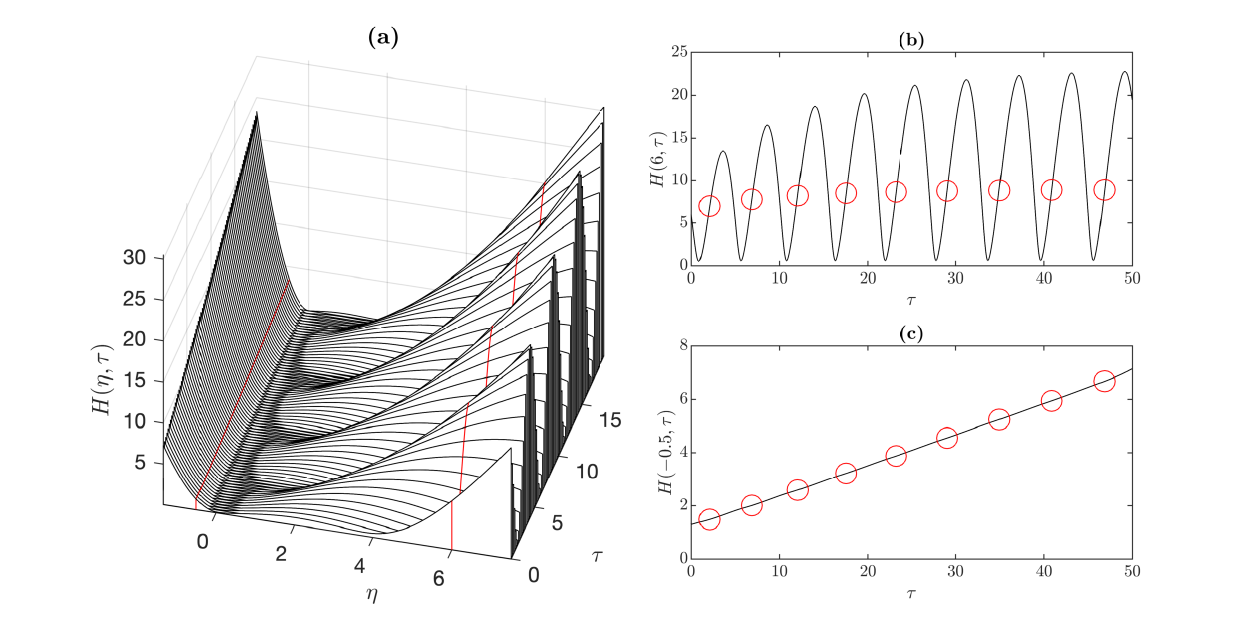}
\caption{Numerical results to \eqref{eq:sMPDE}--\eqref{eq:sMPDE_BC}, highlighting the time-dependent oscillatory behavior of the scaled thin film profile. (a) The evolution of the profile $H(\eta, \tau)$ over a subinterval of the scaled time variable $\tau$. (b) The evolution of $H$ at a specific choice of the similarity variable $\eta = 6$ to the right of the assumed rupture point near $\eta = 0$, demonstrating the oscillatory behavior in this region. (c) The evolution of $=H$ at $\eta = -0.5$, to the left of the rupture point, demonstrating the monotonically increasing slope in this region.  The times indicated by circles in subfigures (b,c) are points where the profile $H$ has a local maximum at $\eta=6$, which are used to calculate the droplet sizes in Fig.~\ref{fig:bumps}. \label{fig:combined_profiles}}
\end{figure*}

In Fig.~\ref{fig:combined_profiles} we show the solution behavior up to a scaled time of $\tau=50$.  The solution profiles are highly asymmetric in $\eta$, with the profile acting in a highly oscillatory fashion for $\eta > 0$, and in a monotonically steepening fashion for $\eta < 0$.  To demonstrate this behavior, we plot the value of $H$ as $\tau$ increases at two fixed values of $\eta$: $\eta = 6 >0$ (Fig.~\ref{fig:combined_profiles}b), and $\eta = -0.5 < 0$ (Fig.~\ref{fig:combined_profiles}c).  We also note that although small, the minimum scaled thickness $H_{\min}(\tau) = \min_{\eta} H(\eta,\tau)$ remains finite (as depicted in Fig.~\ref{fig:marangoni_loglog}b).

However, the solution has not settled down to a periodic orbit in the $(\eta, \tau)$ coordinate system, as can be most clearly seen in the monotonic, seemingly linear behavior of $H$ in $\tau$ when $\eta < 0$ (Fig.~\ref{fig:combined_profiles}b).  If this monotonic increase continues, it indicates that the slope of the scaled profile becomes unbounded in the limit $\tau\to\infty$.  Solution behavior with this property has not previously been observed in the thin film models in which discrete self-similarity arises from the presence of a periodic orbit in rescaled coordinates\,\cite{dallaston_discrete_2018}.


We also use the rescaled solution to extend the observed behavior of the minimum thickness in Fig.~\ref{fig:marangoni_loglog}.  The solid line in this figure represents the minimum thickness found using the rescaled method described in this section, translated back into the original variables.  This plot emphasizes that although not observable for the times achieved when solving \eqref{eq:MPDE} in the original, unscaled coordinate system, for much longer times the minimum thickness asymptotically approaches a regime in which it oscillates around the power law $t^{-1}$.  The oscillations around this power law are connected with the formation of discrete undulations for $\eta>0$ observed in Fig.~\ref{fig:combined_profiles}.

\begin{figure}
\centering
\includegraphics[scale=1]{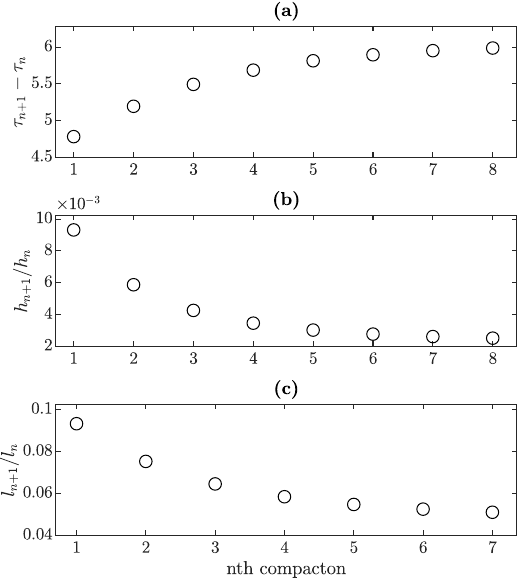}
\caption{ A summary of the spatiotemporal information of numerical solutions to \eqref{eq:sMPDE}--\eqref{eq:sMPDE_BC}. (a) The measured period of the scaled solution. (b, c) The height and the length ratios, respectively, between successive droplets, calculated using \eqref{eq:heightRatio}, \eqref{eq:lengthRatio}. \label{fig:bumps}}
\end{figure}

A meaningful solution property to estimate is the size of successive satellite droplets produced by the long-time dynamics.  In order to estimate the height of and distance between successive droplets, we use the scaled times $\tau_n$ when the local spatial maximum in the solution passes the point $\eta = \eta_0 = 6$ (as in Fig.~\ref{fig:combined_profiles}), which is sufficiently far from zero to represent the far field, but not so close to the boundary to be affected by the imposed boundary condition.  Returning to the original coordinate system $h = h(x,t)$, we estimate the height of the $n$th droplet $h_n$ and distance between the $n$th and $(n+1)$th droplet $l_n$ to be
\begin{equation*}
h_n = \e^{-\alpha \tau_n} H(\eta_0, \tau_n), \quad l_n = \left(\e^{-\beta \tau_n} -  \e^{-\beta\tau_{n+1}}\right)\eta_0,
\end{equation*}
while taking the ratio of these quantities results in
\begin{align}
\frac{h_{n+1}}{h_n} \approx \e^{-\alpha\left(\tau_{n+1} - \tau_n\right)} \frac{H\left(\eta_0, \tau_{n+1}\right)}{H\left(\eta_0,\tau_n\right)}, \label{eq:heightRatio} \\
\frac{l_{n+1}}{l_n} \approx \frac{\e^{-\beta \tau_{n+1}} - \e^{-\beta \tau_{n+2}}}{\e^{-\beta \tau_n} - \e^{-\beta \tau_{n+1}}}. \label{eq:lengthRatio}
\end{align}
If $H(\eta,\tau)$ were exactly periodic with period $\tau_{n+1} - \tau_n = T$, then the ratios $h_{n+1}/h_n = \e^{-\alpha T}$ and $l_{n+1}/l_n = 1-\e^{-\beta T} $ would be independent of $n$, and satellite droplets would be of exactly geometrically decreasing size.  Given our solution plotted in Fig.~\ref{fig:combined_profiles} is not exactly periodic, we do not expect this to be exactly the case.  In Fig.~\ref{fig:bumps}, we plot the apparent periods $\tau_{n+1}-\tau_n$, as well as the ratios of successive heights and lengths for our numerical solution, found using \eqref{eq:heightRatio} and \eqref{eq:lengthRatio}. This amounts to taking the respective height and length ratios of the shapes formed in Fig. \ref{fig:combined_profiles}c, after converting their dimensions back to the original unscaled $(x,t)$ coordinates. Over the time interval calculated, the period is an increasing function of $n$, and the ratios are a decreasing function of $n$.  We do observe that these values are slowly approaching constant values
\[
\tau_{n+1} - \tau_n \to 6, \qquad \frac{h_{n+1}}{h_n} \to 0.002, \qquad \frac{l_{n_1}}{l_n} \to 0.05 
\]
as $\tau$ becomes large, although they have not done so by the end of our simulation ($\tau = 50$).  As this time corresponds to an unscaled time of $t\approx \e^{50} \approx 10^{21}$, this long-time regime where satellite are geometrically decreasing in scale. This is practically unachievable in unscaled numerical simulations (and any real-life situation).

\section{Discussion}

In this paper, we have numerically computed the solution to the thin film equation \eqref{eq:MPDE}, modelling the effects of surface tension and thermocapillarity, for times greatly exceeding what has previously been achieved.  This long-time solution was found by solving the equation in a dynamically rescaled coordinate system \eqref{eq:scale} that focuses on the region where the film thickness vanishes.  This coordinate system is related to similarity analysis in that a similarity solution to the original equation represents a steady state of the rescaled problem \eqref{eq:sMPDE}, and that a periodic solution of the rescaled equation represents a discretely self-similar solution to the original problem, as described in Ref. \cite{dallaston_discrete_2018} for similar models.  However, no assumption of self-similarity is made when solving \eqref{eq:sMPDE} as an initial-value problem, allowing observation of other potential phenomena.\\

The major issues introduced by the rescaled problem is the artificial instability introduced by the translational invariance of the original problem, and the choice of far field boundary conditions \eqref{eq:sMPDE_BC}.  The instability was resolved by choosing an initial condition from the original problem at a sufficiently late time, using a bisection method to determine the appropriate translation to allow the rescaled simulation to run as long as possible.\\

The previous method by Shklyaev et al. \cite{shklyaev_superexponential_2010} provided results by numerically solving up to a point in time, truncating the computational window, applying new boundary conditions, interpolating onto a new mesh, and resume the solver for up to $t = 8000$, at which time four satellite droplets (referred to as compactons) were generated.  In order to extend their solution to later times, they had to artificially restrict the domain to a subset of the original domain, and set pinning conditions on the new subdomain.  We believe that solving the dynamically rescaled problem \eqref{eq:MPDE} is a more natural way to increase resolution in the region of interest.\\

While our simulated results for equation \eqref{eq:sMPDE} oscillate in exactly the way expected to produce an infinite cascade of droplets of successively decreasing size, it is quite clear that the solution has not settled down to a strictly periodic solution in the logarithmic time variable $\tau$.  While the period of oscillations $\tau_{n+1}-\tau_n$ and ratios of length and height scales depicted in Fig.~\ref{fig:bumps} do seem to be very slowly approaching fixed values, the monotonic increase in slope for $\eta < 0$ (shown in Fig,~\ref{fig:combined_profiles}b) indicates that solutions are not being attracted to a finite-amplitude periodic orbit in the scaled variables, in the manner observed in \cite{dallaston_discrete_2018} for different exponents in the thin film equation \eqref{eq:Thin_film_eqn}.

One direction of future research is into whether the behavior observed in \eqref{eq:sMPDE} represents the existence of a periodic orbit under a subtly different assumptions.  One possibility is that logarithmic (in time) corrections are required in the power-law scales in the ansatz \eqref{eq:scale}.  This phenomenon is known to occur in some continuously self-similar problems (for example, the curve shortening equation described in \cite{eggers_singularities_2015}, and related second order equations \cite{hocking1972nonlinear}), but the possible combination of this phenomenon with discrete self-similarity has not previously been observed in any system.  Another possibility is that the solution is tending to a periodic orbit that has a singularity to the left of $\eta=0$, hence the ever-increasing slope observed in Fig.~\ref{fig:combined_profiles}b; this has also never previously been observed in similarity solutions, so remains speculative.

\end{document}